\documentstyle[sprocl]{article}

\def\Journal#1#2#3#4{{#1} {\bf #2}, #3 (#4)}
\def\MNRAS{\em MNRAS}
\def\ApJ{\em ApJ}
\def\AJ{\em AJ}

\def\AA{\em A\&A}
\def\Newa{\em New~Astron.}
\def\PhysRevLett{\em Phys.~Rev.~Lett.}

\def\vcirc{v_{\rm circ}}
\def\lcdm{{$\Lambda$CDM}}

\begin{document}

\title{NO NEED FOR DARK MATTER IN GALAXIES?}

\author{N.W. Evans}

\address{Theoretical Physics, Department of Physics, 1 Keble Road,\\
Oxford OX1 3NP, England\\ E-mail: nwe@thphys.ox.ac.uk}
 

\maketitle\abstracts{Unhappily, there has been a maelstrom of problems
for dark matter theories over the last few years and many serious
difficulties still have no resolution in sight. This article reviews
the evidence for dark matter in galaxies. Judged on the data from
galactic scales alone, the case for dark matter is weak.
Non-Newtonian theories of gravity have their own problems, but not on
galactic scales.}

\section{Introduction}

The best evidence for mass discrepancies in galaxies comes from
rotation curves of neutral hydrogen. However, in the case of our own
Galaxy, the rotation curve cannot be traced beyond $\sim 20$ kpc. It
is the kinematics of stellar tracers of the distant halo -- namely the
bound dwarf galaxy satellites and distant globular clusters -- that
provide the mass estimates. The most recent analysis~\cite{mw} found
the total dynamical mass of the Galaxy is $\sim 2 \times 10^{12}
M_{\odot}$. By contrast, the mass of the stellar disk is $\sim 6
\times 10^{10} M_\odot $ and of the stellar bulge and spheroid is
$\sim 3 \times 10^{10} M_\odot $. There are only two
alternatives. The first is that the Galaxy is largely composed of dark
matter. This is matter whose existence is inferred solely from its
gravitational effects.  The second is that Newtonian gravity is
incorrect and needs modification. There is no hard evidence that
conventional theories of gravity are correct on scales greater than at
most a parsec.  But, to many astronomers, the second option is
unattractive, as the hard-won understanding of large-scale structure
formation would then be jeopardised~\cite{rees}.

If dark matter exists, what is it?  Microlensing surveys~\cite{alcock}
and {\em Hubble Space Telescope} pencil-beam searches~\cite{graff}
strongly constrain the mass in dim stars and brown dwarfs. Suppose we
rashly assume that all the lenses towards the Magellanic Clouds lie in
the Galactic halo~\cite{sahu}, then no more than $\sim 9 \times
10^{10} M_\odot$ of the mass within 50 kpc lies in such objects. Yet,
the dynamically inferred mass is roughly five times greater than this.
So, there is firm evidence that faint or failed stars do not make up
most of the Galactic halo.  In contrast, non-baryonic particle dark
matter is an essential component of theories of structure
formation. First, the total dynamical mass density in the Universe
exceeds that permitted for baryonic dark matter by big bang
nucleosynthesis, and second, the growth of structure requires
pre-existing perturbations in the non-baryonic component at the time
of recombination~\cite{kt}.  Cold dark matter theories (CDM) or their
variants with a cosmological constant (\lcdm) give a fine description
of large-scale structure~\cite{frenk}, but there remain stubborn
problems on the scales of kiloparsecs. {\it The recent years have seen
increases in the quality, the quantity and the severity of these
problems.} Some astronomers have been led to question the very
existence of dark matter~\cite{jerry}, while others have been led to
imbue the dark matter with additional, exotic
properties~\cite{dands}. It is a measure of the seriousness of the
situation that such drastic hypotheses are being put forward.

\section{The Problems}

A galaxy is described as having a ``maximum disk'' when the dark
matter contribution to the central attraction in the inner regions is
negligible compared to that of the luminous disk and bulge or bar.
The haloes built up by hierarchical merging in dark matter cosmogonies
(such as the Navarro-Frenk-White or NFW models) are cusped and
dominated by dark matter at the very center~\cite{nfw}.  However,
there is overwhelming evidence that the disks of large spiral galaxies
are close to maximum and the haloes make only a small contribution to
the rotation curve in the inner parts.  For the dark-matter dominated
low surface brightness and dwarf galaxies, matters are less clear-cut,
but even here, the balance of the evidence is against cusped haloes.

\subsection{The Microlensing Optical Depth to the Galactic Center}

For example, there are extremely high microlensing optical depths
towards Baade's Window~\cite{macho}.  The optical depth to the red
clump stars is $3.9 \times 10^{-6}$. This is a crucial measurement as
the red clump stars are bright (so the efficiency of the survey is
high) and are known to reside in the Galactic bulge (so that
contamination by background sources is unlikely).  Even barred Galaxy
models have real difficulty in generating the measurements without
violating constraints on the Galactic force-field, while axisymmetric
models are ruled out. The power of this argument is that it does not
depend on the details of the modelling. It follows from computing the
maximum contribution to the microlensing optical depth, consistent
with the rotation curve, made by circular or elliptical rings of
matter~\cite{jjb}. Therefore, almost all the matter along the lines of
sight to the Galactic bulge must be capable of causing microlensing
and cannot be particle dark matter. This argument immediately rules
out cusped profiles (like the NFW model) as realistic representations
of the present-day structure of the Galaxy halo. They give values of
the optical depth to the red clump stars that are seriously awry.

\subsection{Hydrodynamical Modelling of the Galactic Bar}

The other strong, albeit more model dependent, piece of evidence that
the Galaxy has a maximum disk comes from hydrodynamical and stellar
dynamical modelling of the Galactic bar~\cite{bars}.  First, Weiner \&
Sellwood used a flux-splitting Eulerian grid code to model the gas
flow in a reasonably realistic analytical force field. Second,
Englmaier \& Gerhard used smooth-particle-hydrodynamics to model the
gas flow in the inner Galaxy using a mass distribution derived from
the COBE surface photometry with a constant mass-to-light ratio.  The
studies conclude that to reproduce the terminal velocities of the HI
gas, the bar and the disk must provide almost all the rotational
support within the solar circle. Further evidence is provided by the
stellar velocity dispersions and mean motions available at certain
unobscured windows in the bulge. H\"afner et al. built a stellar
dynamical bar (using a similar mass model and normalisation as
Englmaier \& Gerhard) and showed it reproduced essentially all of the
available stellar kinematical data in the inner Galaxy.

\subsection{Stability of Unbarred Disk Galaxies}

It is the problem of galactic stability that led to the introduction
of dark haloes in the first place.  Many simple models of
self-gravitating disks disrupt to form bars on a dynamical time scale.
Ostriker \& Peebles~\cite{op} argued that unbarred galaxies are stable
because they contained a large fraction of mass in a dynamically hot
component which was unable to participate in collective instabilities.
Although they never said this, their paper has sometimes been
misrepresented as stating that maximum disks are unstable. This
misunderstanding persists in some papers on galaxy formation to this
very day.

Toomre~\cite{toomre} showed that disks could be stabilized by making
their centers impenetrable to density waves.  One way to achieve this
is for the circular speed to remain high towards the center, which
forces an inner Lindblad resonance (ILR) that cuts the feedback loop
to the swing amplifier.  {\it Even a fully self-gravitating disk can
be stabilized against bisymmetric modes.}  Toomre's prediction was
almost immediately contested by numerical simulations~\cite{eln},
which found no evidence for stabilization by hard centers. This
contradiction was resolved by the realisation that large-amplitude
disturbances in highly responsive disks could saturate the ILR,
trapping particles in a large-scale bar similar to that which would
have formed without the dense center~\cite{jas}.  The recent years
have seen a number of examples of robustly bar-stable galaxy models
having cool, maximal disks with dense centers~\cite{many}.  For
example, Sellwood \& Evans have constructed a completely stable and
reasonably realistic galaxy model in which the disk provides most of
the central attraction in the inner parts. This model has a
quasi-exponential disk and an almost flat rotation curve that arises
from a combination of the massive disk, a central bulge, together with
a halo having very low central density. They showed, using both linear
stability theory and numerical simulations, that it has no global
instabilities whatsoever.

What stabilizes real unbarred disk galaxies: hard centers or dark
haloes?  High-resolution kinematic observations of spiral galaxies
have uncovered high orbital speeds close to their
centers~\cite{sofue}.  Galaxies with gently rising rotation curves at
the center all have low luminosity.  Such galaxies are required to
have large dark matter fractions to suppress bar-forming
instabilities.  However, almost every galaxy with a circular speed in
excess of 150 km s$^{-1}$ has a steep inner rise in the rotation curve
and must be bar-stable whatever its dark matter content.  Bright
unbarred disk galaxies are stabilized by their hard centers.

\subsection{Pattern Speeds of Bars}

Bars in galaxy models with cusped dark halos all experience strong
drag from dynamical friction~\cite{victor}. The drag causes the bar to
slow down and drives the corotation point out to distances well beyond
the bar's optical edge. This process acts swiftly, on the timescale of
a few bar rotation periods. If dark matter dominates the central parts
of barred galaxies, then bars are expected to be slow.  Pattern speeds
have now been measured for a handful of barred galaxies~\cite{mm}. The
sample is small, but the results are consistent and confirm the
theoretical expectation that bars in real galaxies are fast
rotators. The corotation point typically lies just beyond the bar's
optical edge.  Bars can maintain such high pattern speeds only if the
disk provides most of the central attraction in the inner regions.

\subsection{Rotation Curves}

All the above evidence depends on dynamics and is much more reliable
than the evidence based on rotation curves. Unsurprisingly, there is
inevitable degeneracy in decomposing galaxy rotation curves into
contributions from the disk, bulge and halo, as well as assessing the
importance of experimental effects like beam-smearing. Consequently,
this area has been beset by controversy.

For bright galaxies, maximum disks are favored by the fact that
luminous matter alone accounts for the overall shape of the rotation
curve in the inner parts. This is a point of view originally advocated
by Kalnajs in 1983.  For example, an extensive recent survey of 74
southern spirals found that $\sim 75 \%$ are well-fitted by a
mass-traces-light model for the inner parts~\cite{palunas}.  For dwarf
and low surface brightness (LSB) galaxies, dark matter dominates
everywhere.  Early claims that the rotation curves are inconsistent
with cusped haloes have given way to an acceptance that the available
data may not be good enough to decide the issue. The HI rotation
curves of the LSB disks, and probably the dwarf spirals as well, are
strongly affected by beam-smearing and are consistent with both cusps
and cores~\cite{vdb}. Even so, for the dwarf spiral NGC 5585, the
highest quality HI and H$\alpha$ data do appear to show that the halo
density profile rises to a finite and uncusped value~\cite{bo}.  There
is also a sample of five LSB galaxies for which high resolution
H$\alpha$ rotation curves are available~\cite{madore}. This study
demonstrates that there is a tendency for the HI data to underestimate
the inner slopes of the rotation curves.  Two galaxies in the sample
have much steeper inner slopes in H$\alpha$ than HI, two are mildly
steeper, while one (F568-3) is unchanged.

A fair summary is that the data favor the absence of cusps, but the
evidence is far from conclusive. Only when the HI data is complemented
by H$\alpha$ observations can robust conclusions be drawn. Then, there
do appear to be examples of galaxies -- like the LSB galaxy F568-3 and
the dwarf spiral NGC 5585 -- which are only consistent with cores. On
the other hand, there is no clear-cut example of a galaxy whose halo
requires a cusped profile.

\section{Conspiracies and Catastrophes}

Once it is accepted that bright spiral galaxies are dominated by
baryons in the inner parts, then dark matter theories become engulfed
by two serious problems -- the disk-halo and the surface brightness
conspiracies.  No convincing solution of these difficulties have ever
been proposed. High resolution N-body simulations have also uncovered
a number of problems with galaxy formation in cold dark matter (CDM)
cosmogonies, the most serious of which are the angular momentum and
satellite catastrophes. (Other problems are listed in Sellwood
\& Kosowsky~\cite{jerry}). 

\subsection{The Disk-Halo Conspiracy}

There is no feature in the galaxy rotation curve as the dominant
source of the gravity field changes from luminous to dark~\cite{cb}.
In its original formulation, CDM removed this difficulty, as the dark
matter dominated everywhere, even at the very centers of bright
galaxies. Baryonic disks formed in the pre-existing potential wells of
dark haloes. However, this is no longer tenable as the evidence that
baryons dominate the central parts of bright galaxies is too strong.
It is very hard to understand why the circular velocity from luminous
matter at the center should be so similar to that from dark matter at
large radii.  If the dark matter and the luminous matter are
unrelated, the resolution of this difficulty requires fine tuning.

\subsection{The Surface Brightness Conspiracy}

On dimensional grounds, we expect that $\vcirc^2 \sim G M /L$, where
$\vcirc$ is the circular speed, $M$ is the total mass and $L$ is the
characteristic size of a galaxy. So, we expect a low surface
brightness galaxy to have a lower circular velocity than a high
surface brightness of the same mass. This is not the case. We observe
similar circular speeds in all galaxies of a given luminosity, no
matter how widely dispersed the luminous material.  In other words,
extremely low surface brightness galaxies lie on the same Tully-Fisher
relation as that derived for high surface brightness
galaxies~\cite{Z}.  This requires the dark matter fraction to rise as
the luminous surface density declines. This is not easy to arrange
without fine tuning, as the baryons dominate the mass of the inner
parts of bright galaxies.

\subsection{The Angular Momentum and Satellite Catastrophes}

Any hierarchical merging cosmogony suffers from the angular momentum
catastrophe. As they cool, the baryons lose angular momentum to the
halo making disks that are much too small~\cite{KG}. The predicted
scalelengths of disk are at least a factor of $\sim 5$ less than
observed.  CDM cosmogonies also suffer from the satellite
catastrophe. High resolution N-body simulations have revealed large
numbers of sub-clumps within large dark matter haloes, typically
orders of magnitude more than are observed as satellite galaxies of
(say) the Milky Way or M31~\cite{klypin}. Both these catastrophes are
a consequence of the persistence of sub-structure. Warm dark matter
cosmogonies can partly evade the second problem, but not the first.

Processes such as feedback from star formation, explosions, ejection
from gaseous bars and massive galactic winds have all been invoked to
solve these problems and to remove the central cusp of the dark halo
as well. For example, one suggestion is that galaxies first absorbed
and then ejected a mass of baryons that is comparable to their current
baryonic mass~\cite{bsilk}. The energy and angular momentum given up
by the ejected baryons modify the structure of the dark halo, erasing
the central cusp and removing substructure. Galactic disks are formed
from material at the periphery of the volume from which the galaxy's
dark matter was drawn. On account of its large galactocentric radius,
this material had more angular momentum than the disk it finally
built.

Only simulations will tell us whether feedback processes can solve
these catastrophes. At present, the simulations do suffer from
resolution difficulties and lack some of the essential physics. At
least as judged from the available simulations~\cite{mf}, it looks
likely that feedback can solve the satellite catastrophe, but it looks
unlikely that it can actually erase steep cusps.

\section{Discussion}

There is no observational evidence that requires dark haloes to have
central cusps.  And there is a lot of hard-to-circumvent dynamical
evidence that they do not.  It is worth stressing that the really
damaging evidence against cusps comes from the dynamics and not from
the rotation curves.  The alternative to dark matter -- namely
non-standard theories of gravity -- has been much less well
explored. Nonetheless, it is striking that a theory like
MOND~\cite{ssm} resolves the disk-halo conspiracy and the surface
brightness conspiracy at the cost of introducing a new fundamental
scale.  (In fact, it was a successful prediction of MOND that low
surface brightness galaxies obey the same Tully-Fisher relation as
high surface brightness galaxies). There is no theory of cosmology or
galaxy formation available in MOND, so nothing can be said about the
angular momentum or the satellite catastrophes.  {\it Judged on the
evidence from galactic scales alone, the case for dark matter is weak
and MOND is the better theory}.  Non-Newtonian theories of gravity
have problems of their own (such as lack of covariance), but not on
galactic scales.

A characteristic of most ultimately successful theories is that they
work best in the r\'egimes where the data are best understood, and
work least well in the r\'egimes where the data are poorest.  CDM (and
its variants) do not have this property.  It is on galactic scales
that distances are best known, samples are most complete and the data
are best understood and here CDM has a number of stubborn problems. By
contrast, the great triumph of CDM is in the reproduction of the
large-scale distribution of galaxies.

\end{document}